\documentclass[twocolumn,showpacs,aps,prl,superscriptaddress,floatfix]{revtex4}
\usepackage{graphicx}
\usepackage{dcolumn}
\usepackage{bm}
\usepackage{amsmath}
\usepackage{color}

\begin{document}

\preprint{}

\title{Order by Disorder Spin Wave Gap in the XY Pyrochlore Magnet Er$_{2}$Ti$_{2}$O$_{7}$}

\author{K.A. Ross}
\affiliation{Institute for Quantum Matter and Department of Physics and Astronomy, Johns Hopkins University, Baltimore, Maryland 21218, USA}
\affiliation{NIST Center for Neutron Research, National Institute of Standards and Technology, Gaithersburg, Maryland 20899, USA}
%

%

\author{Y.Qiu} 
\affiliation{NIST Center for Neutron Research, National Institute of Standards and Technology, Gaithersburg, Maryland 20899, USA}
\affiliation{Department~of~Materials~Science~and~Engineering,~University~of~Maryland,~College~Park,~Maryland~20742,~USA}
%
\author{J.R.D. Copley} 
\affiliation{NIST Center for Neutron Research, National Institute of Standards and Technology, Gaithersburg, Maryland 20899, USA}

\author{H.A. Dabkowska} 
\affiliation{Brockhouse Institute for Materials Research, McMaster University, Hamilton, Ontario, L8S 4M1, Canada}

\author{B.D. Gaulin} 
\affiliation{Department of Physics and Astronomy, McMaster University,
Hamilton, Ontario, L8S 4M1, Canada}
\affiliation{Brockhouse Institute for Materials Research, McMaster University, Hamilton, Ontario, L8S 4M1, Canada}
\affiliation{Canadian Institute for Advanced Research, 180 Dundas St.\ W.,Toronto, Ontario, M5G 1Z8, Canada}


\bibliographystyle{prsty}

\begin{abstract}
The recent determination of a robust spin Hamiltonian for the anti-ferromagnetic XY pyrochlore Er$_{2}$Ti$_{2}$O$_{7}$ reveals a most convincing case of the ``order by quantum disorder'' (ObQD) mechanism for ground state selection.   This mechanism relies on quantum fluctuations to remove an accidental symmetry of the magnetic ground state, and selects a particular ordered spin structure below T$_N$=1.2K.  The removal of the continuous degeneracy results in an energy gap in the spectrum of spin wave excitations, long wavelength pseudo-Goldstone modes.  We have measured the ObQD spin wave gap at a zone center in Er$_{2}$Ti$_{2}$O$_{7}$, using low incident energy neutrons and the time-of-flight inelastic scattering method.  We report a gap of $\Delta$=0.053 $\pm$ 0.006 meV, which is consistent with upper bounds placed on it from heat capacity measurements and roughly consistent with theoretical estimate of $\sim$ 0.02 meV, further validating the spin Hamiltonian that led to that prediction.  The gap is observed to vary with square of the order parameter, and goes to zero for $T \sim T_{N}$.
\end{abstract}


\maketitle



Geometrically frustrated magnetic materials consist of interacting localized magnetic moments arranged on crystalline architectures for which the anisotropies and interactions of the moments are incompatible with a simple ordered state.  Frustration suppresses conventional phase transitions to long range magnetic order, allowing the study of strongly correlated and fluctuating spins at temperature scales much lower than the interaction energy \cite{lacroix2011introduction}.  Among the exotic ground states which can result are quantum entangled states, such as the much discussed $U(1)$ quantum spin liquid \cite{hermele2004pyrochlore}.   However, many frustrated systems do find their way to long range order via weak energetic terms in the Hamiltonian, ordinarily insignificant in a non-frustrated system.  This provides many possibilities for the ground states of such magnets.   Commonly encountered terms beyond nearest neighbor exchange which may be relevant in the effective spin Hamiltonian include long range dipolar interactions, next nearest neighbor exchange, and spin-lattice coupling terms.

\begin{figure*}[!htb]  
\centering
\includegraphics[ width=14cm]{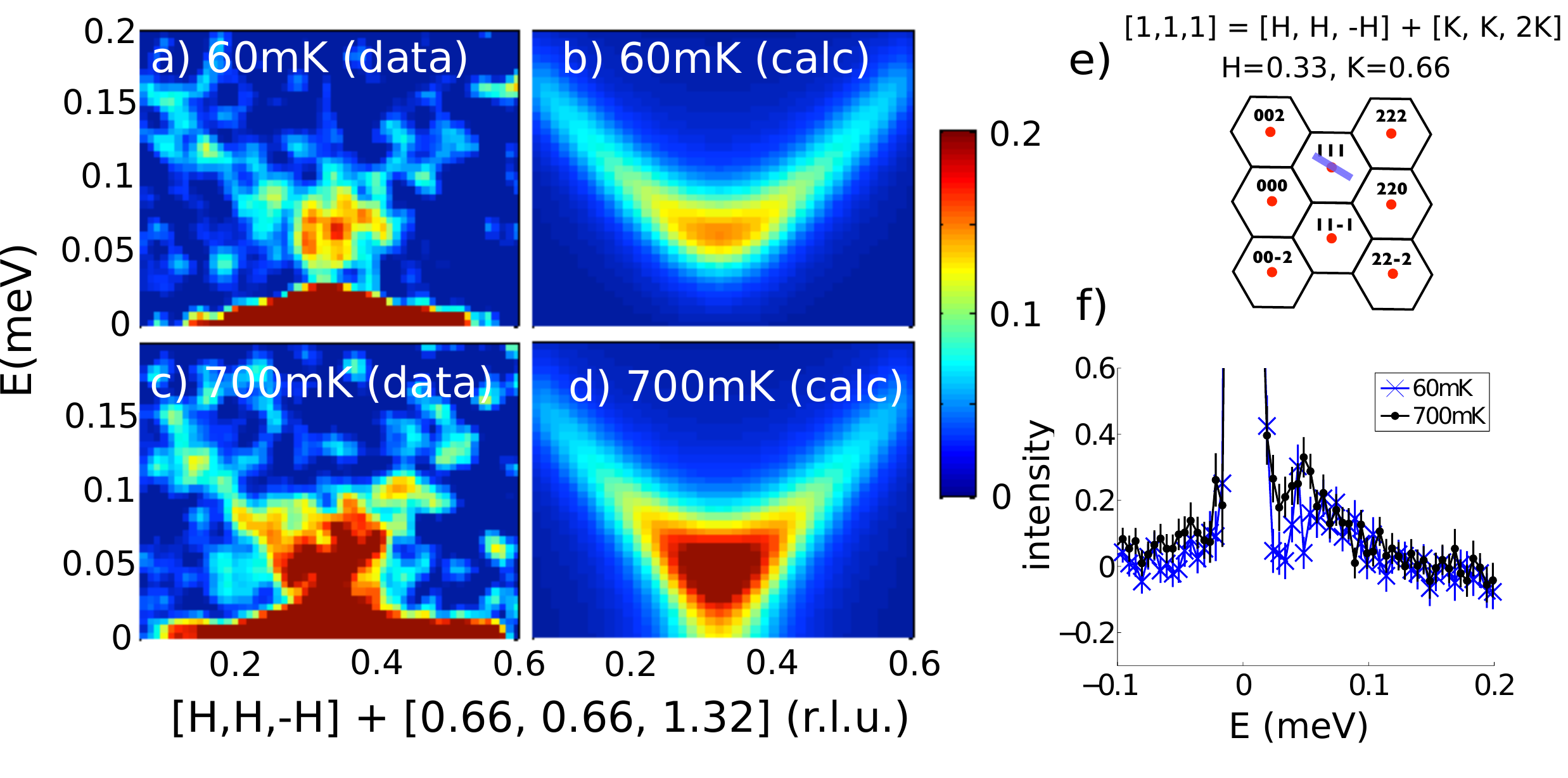}
\caption{ a) through d): color contour maps of the INS spectrum at the bottom of the spin wave dispersion near the (1,1,1) Bragg peak, along the [1,1,-1]  direction ([H,H,-H] + [K,K,2K], with 0.58 $\le$ K $\le$ 0.74 r.l.u.), as shown in a representation of the reciprocal plane (e). Panels a) and c) show the measured $S(\vec{Q},\omega)$ at $T=$ 60mK and 700mK, while panels b) and d) show fits to these data using Eq. \ref{eqn:scalc}, (see text). f) Energy cuts, binning over 0.30 $\le$ H $\le$ 0.36 r.l.u.  Errorbars represent one standard deviation.  Note that the factor $k_f/k_i$ which appears in the INS cross section is removed in all data presented herein.}
\label{fig:fig1}
\end{figure*}

``Order-by-Disorder'' (ObD) is an appealing, yet elusive, mechanism to drive order in frustrated systems \cite{villain1980, shender1982antiferromagnetic, henley1989ordering, lacroix2011introduction}.  Generally, ObD breaks an ``accidental'' continuous degeneracy, i.e. one which is not supported by the symmetry of the Hamiltonian, by accessing low energy fluctuations around these ground states; the favoured states are those with higher densities of low energy modes.  Thus, ObD selects a ground state by entropic rather than energetic means.  In magnetic systems the selection occurs via low energy spin fluctuations, either thermally or quantum mechanically driven, the latter case being called Order-by-Quantum-Disorder (ObQD).  The same mechanism can quite generally apply to other types of ensembles, such as ultra-cold atoms and lattice boson models \cite{barnett2012order, turner2007nematic, melko2005supersolid}.  While there is no lack of theoretical \emph{models} which definitively display ObD \cite{villain1980, shender1982antiferromagnetic, henley1989ordering, reimers1993order, henley1995ground, champion2003, yildirim1994spin, henley1994selection}, the number of real systems which display ObD are few  \cite{gukasov1988quantum, kim1999ordering2} and even in those cases, their ground state selection usually cannot be \emph{conclusively} ascribed to ObD  (see supplementary material in Ref. \onlinecite{savary2012order}).  The pyrochlore magnet Er$_2$Ti$_2$O$_7$ is a very rare example of a magnetic material that can be compellingly shown to display ObQD, thanks to a detailed knowledge of its effective spin Hamiltonian and the unusually well-protected accidental degeneracy it supports \cite{zhitomirsky2012quantum, savary2012order}.  Here, we report direct measurements of the defining characteristic of the excitation spectrum which further supports ObQD in Er$_2$Ti$_2$O$_7$; a  small spin wave gap that is induced by quantum fluctuations.



ObQD introduces an anisotropy into the Hamiltonian through the excitation of spin wave modes supported by the ordered ground state.  This anisotropy manifests itself as an energy gap to the long wavelength excitations near $\vec{Q}= (0,0,0)$ (and related zone centers).  These excitations would otherwise be Goldstone modes within the continuously degenerate manifold.  However, unlike crystalline anisotropy, the ObQD anisotropy is removed as one exits the ordered state.  A continuous phase transition out of a ground state selected by ObQD is then expected to be accompanied by a continuous closing of the gap.  


Being the canonical example of geometric frustration in 3D, the pyrochlore lattice, an array of corner-sharing tetrahedra, often supports large ground state degeneracies.  Spins interacting via Heisenberg near neighbor antiferromagnetic (AF) exchange on the pyrochlore lattice are well known to display a macroscopic degeneracy of \emph{disordered} ground states \cite{reimers1992absence, moessner1998properties}.   A simple modification to the near-neighbor AF pyrochlore model is the introduction of XY anisotropy.  In real pyrochlore materials, the anisotropy is local, resulting from the D$_{3d}$ point symmetry at the $R$-site of the crystal lattice in materials such as $R_2$Ti$_2$O$_7$ (with $R$ being a trivalent rare earth ion).  For local XY anisotropy and AF exchange interactions, it was discovered that a large \cite{bramwell1994order} continuous \cite{champion2003, champion2004}, manifold of \emph{ordered} states arises, but this degeneracy is lifted by ObD \cite{champion2003, champion2004, stasiak2011order}, such that a small discrete set of ordered spin configurations is selected.

Er$_{2}$Ti$_{2}$O$_{7}$ is the only known manifestation of an AF XY pyrochlore magnet.  Large single crystals of Er$_{2}$Ti$_{2}$O$_{7}$ can be produced via the optical floating zone (OFZ) method,  allowing a comprehensive study of its magnetic properties including anisotropic effects.   An estimate of the overall exchange energy comes from Curie-Weiss fits to the susceptibility, giving $\theta_{CW}$= -13 K at high temperatures \cite{bramwell2000bulk, dasgupta2006crystal}.  Meanwhile, the crystal field Hamiltonian for Er$^{3+}$ in Er$_{2}$Ti$_{2}$O$_{7}$ is known to display a well-isolated Kramers ground state doublet with local XY anisotropy \cite{cao2009, dasgupta2006crystal}.   Er$_{2}$Ti$_{2}$O$_{7}$ is known to order magnetically into the $\psi_2$ basis of the $\Gamma_5$ irreducible representation at T$_N$ = 1.2K \cite{poole2007}, although remnant diffuse magnetic neutron scattering near the Bragg peaks \cite{ruff2008, de2012magnetic} indicates some short ranged spin structure persists.  This diffuse scattering is not fully understood but it is plausible to ascribe it to domain walls separating the six possible domains of the $\psi_2$ ordered state.

While $\psi_2$ order in Er$_{2}$Ti$_{2}$O$_{7}$ is not disputed experimentally, the mechanism leading to it was not fully understood until recently, when an analysis of the full spin wave spectra in the magnetic field-polarized quantum paramagnetic state of Er$_{2}$Ti$_{2}$O$_{7}$ revealed that an anisotropic exchange Hamiltonian with pseudospin-1/2 operators is needed to describe the measured magnetic excitations in Er$_{2}$Ti$_{2}$O$_{7}$ \cite{zhitomirsky2012quantum, savary2012order}. This anisotropic exchange model and related variants is enjoying much current attention  \cite{jointpaper, bonville2013magnetization, wong2013ground, yan2013living}.  The model with parameters as determined in Ref. \onlinecite{savary2012order} was recently studied including the effects of finite temperature, which showed that both thermal and quantum ObD lead to the same ground state selection, i.e. the $\psi_2$ state \cite{oitmaa2013phase}. It has also been argued that $\psi_2$ could be selected energetically through a small admixing of higher crystal field levels \cite{mcclarty2009energetic}, although the magnitude of such an effect has not been rigorously calculated, and indeed by estimates based on the crystal field splitting in Er$_2$Ti$_2$O$_7$, it is expected to be negligible \cite{savary2012order}.  Importantly, Savary \emph{et al} showed that the continuous degeneracy of the lowest energy spin configurations at the mean field level cannot be broken by longer range interactions or spin-lattice coupling, thus cementing ObQD as \emph{the} mechanism leading to the selection of $\psi_2$ order in Er$_{2}$Ti$_{2}$O$_{7}$ \cite{savary2012order}.


 A remaining task is to directly confirm the ObQD scenario through the measurement of the small spin wave gap that is a necessary consequence of it.  Here we report a direct spectroscopic measurement of the spin wave gap using high resolution inelastic neutron scattering (INS) on a single crystal of Er$_{2}$Ti$_{2}$O$_{7}$.  We used the theory from Savary et al, who predicted a gap of approximately 0.02meV \cite{savary2012order}, as well as upper limits of $\lesssim$ 0.05 meV from analysis of low temperature heat capacity measurements \cite{de2012magnetic} and electron spin resonance \cite{sosin2010magnetic} in order to carefully plan these direct measurements of the ObQD spin wave gap.


\begin{figure}[!htb]  
\centering
\includegraphics[ width=8.5cm]{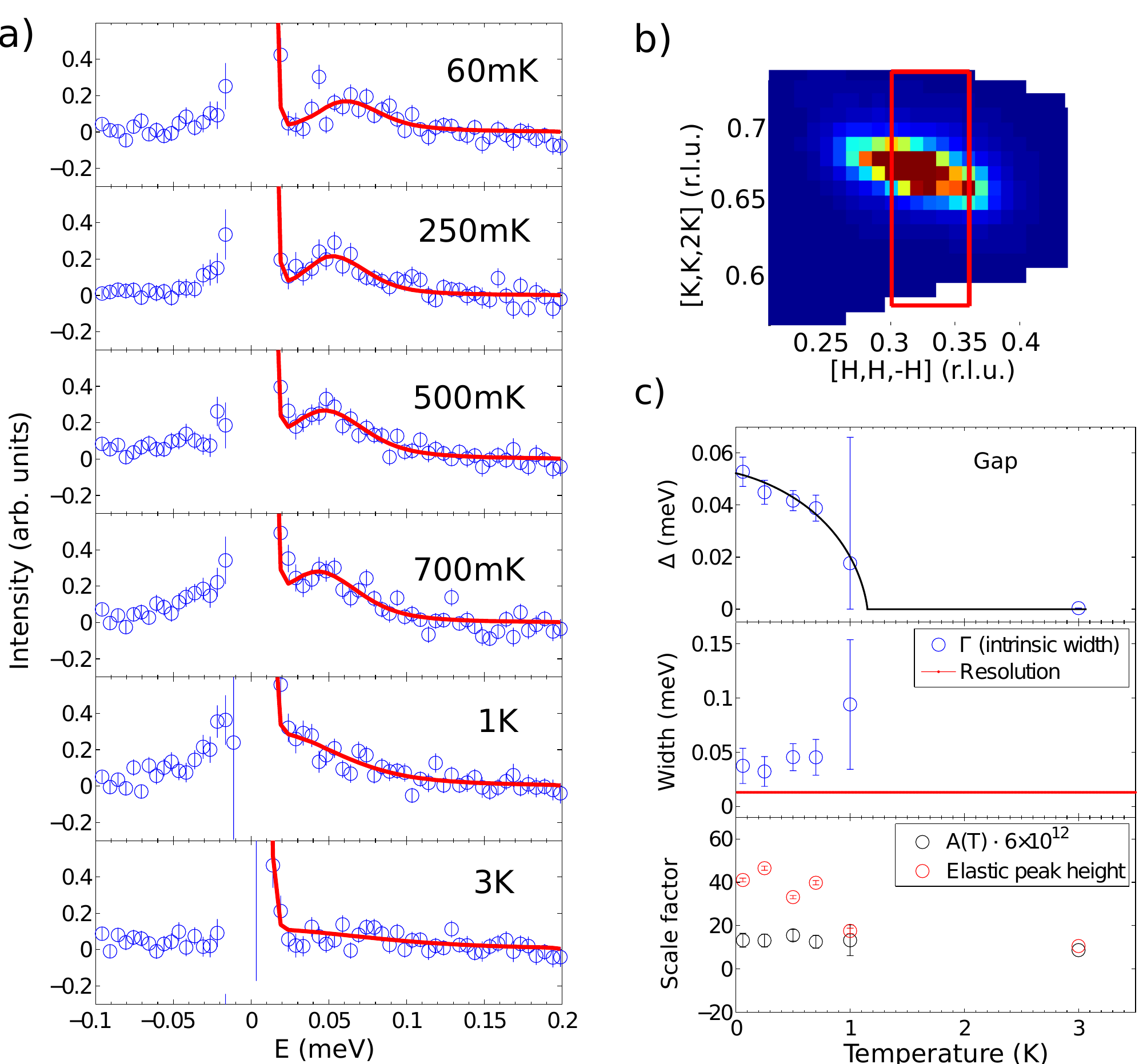}
\caption{ a) Intensity vs. energy cuts through the measured $S(\vec{Q},\omega)$, binning over the region 0.30 $\le$ H $\le$ 0.36, and 0.58 $\le$ K $\le$ 0.74 r.l.u.  b) the binning region overlaid on a map of the elastic scattering ((111) Bragg peak).  The fitting function allows the gap ($\Delta$), intrinsic width ($\Gamma$), and scale factors for both elastic and inelastic intensities ($A$, see Eq. \ref{eqn:scalc}) to vary.  The fitted values are shown in c), where error bars indicate half of the 95\% confidence interval. }
\label{fig:fig1}
\end{figure}

\begin{figure}[!htb]  
\centering
\includegraphics[ width=8.5cm]{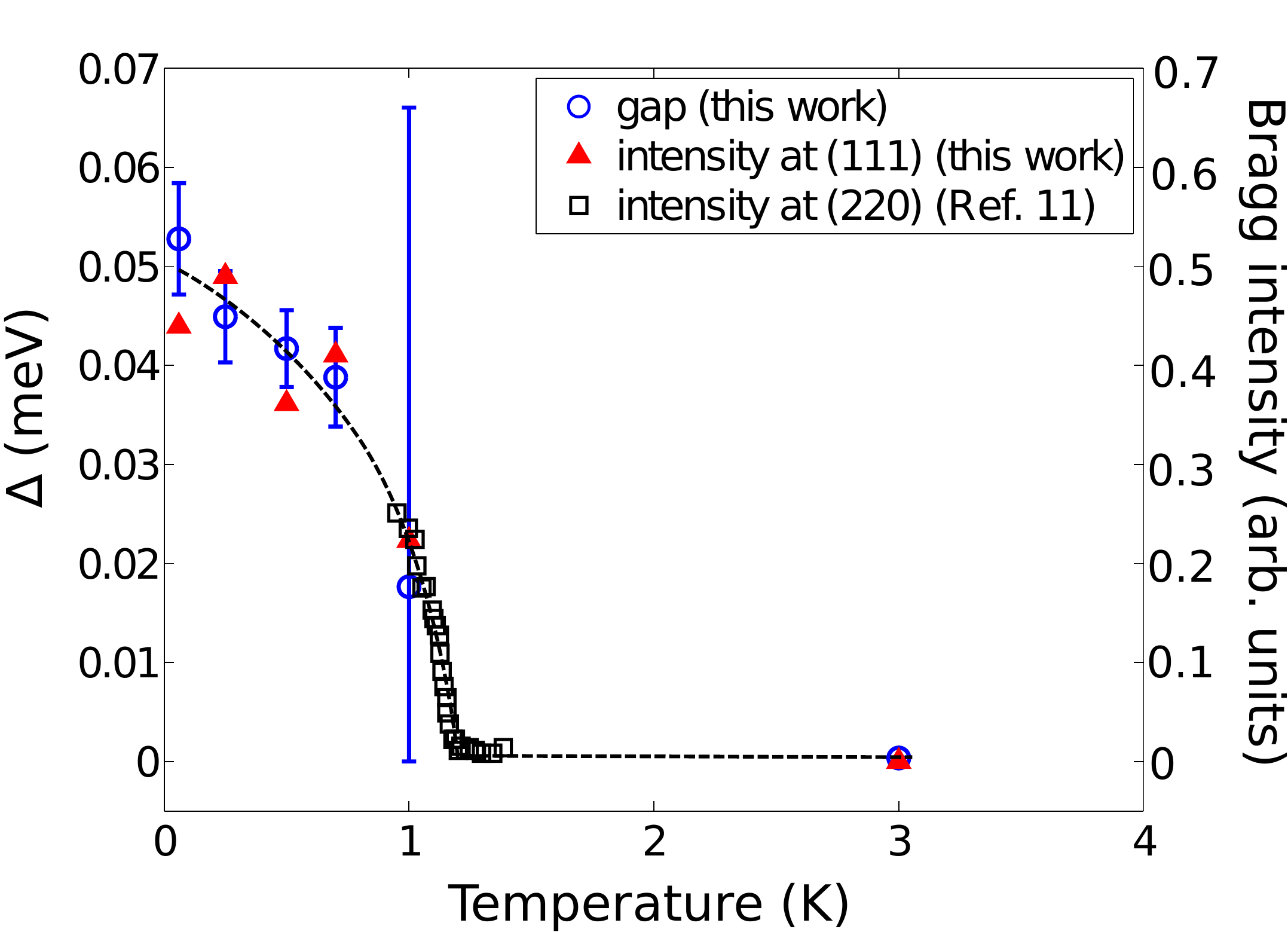}
\caption{ Temperature dependence of the spin wave gap, $\Delta$, overlaid on that of the Bragg peak intensity ($\propto M(\vec{Q})^2$), measured in the present work, as well as that reported in Champion \emph{et al} \cite{champion2003}.  Errorbars on the gap data half of the 95\% confidence interval while those on the Bragg intensity (one standard deviation) are smaller than the markers.}
\label{fig:fig1}
\end{figure}



We studied single-crystalline Er$_{2}$Ti$_{2}$O$_{7}$, grown by the OFZ method \cite{Gardner_growth, dabkowska2010crystal}.  The 7 gram crystal studied here is the same as that previously studied \cite{savary2012order, ruff2008}.  Several other single crystals have also been studied in the literature.  Unlike some other rare-earth titanate materials (notably Yb$_2$Ti$_2$O$_7$ \cite{Yaouanc2011, ross2011dimensional, ross2012lightly} and Tb$_2$Ti$_2$O$_7$ \cite{takatsu2012quantum, taniguchi2013long}), different powders and OFZ growths of Er$_{2}$Ti$_{2}$O$_{7}$ appear to have consistent magnetic properties down to 50mK, as evidenced by the ordering temperature, ordered spin structure, and magnetic specific heat \cite{de2012magnetic}.    

The neutron scattering experiment was carried out at the NIST Center for Neutron Research, using the Disk Chopper Spectrometer.  This time-of-flight instrument produces monochromatic pulses of neutrons from a continuous source using seven choppers in the incident flight path.   We employed the lowest resolution chopper settings, and selected neutrons with incident wavelength of 10 \AA \ ($E_{i}$ = 0.82meV).  This provided an energy resolution of 0.013meV, seven times better than the original INS experiment on Er$_{2}$Ti$_{2}$O$_{7}$ \cite{ruff2008}.  The improved resolution gave a reduced incident neutron flux, 4.6 times lower than the previous study, providing a challenge for the detection of even the strongest inelastic scattering (acoustic spin wave modes). 

The experiment was carried out in a 10T superconducting magnet with dilution insert.  This allowed us to reach well into the ordered state (60mK $\sim$ T$_{N}$/20). At 60 mK the order parameter is approximately saturated and the ObQD gap should be well-defined and at its maximum value.  The application of a 5T magnetic field along the [110] direction was used to lift the magnetic excitations out of the energy window of interest \cite{ruff2008}, allowing a precise determination of the instrumental background.  This background was subtracted in all figures shown here.

%



Figure 1 shows INS data near the (1,1,1) magnetic zone center.  The intensity in a) and c) is shown as a function of energy and along the [1,1,-1] direction in reciprocal space at T=60 mK and 700 mK, respectively.  The trajectory in reciprocal space for these scans is shown in Fig. 1 e).  It is clear that the raw magnetic scattering at T=60 mK in Fig. 1 a) shows a gapped spectrum at (1,1,1), with an approximate gap of $\sim$ 0.05 meV.  On warming up to 700 mK (Fig. 1 c)) the gap has largely closed.  Cuts of this data along the energy direction at T=60 mK and 700 mK and integrating over 0.30 $\le$ H $\le$ 0.36 r.l.u. are shown in Fig. 1 f).

To allow a more quantitative estimate of the ObQD spin wave gap and its temperature dependence, we have fit these data as shown in Figs. 1 c) and d), and in Fig. 2 a).  To fit the measured spin wave dispersion, we require a model for $S(\vec{Q}, \omega)$ around (1,1,1), which we now describe.

The magnetic structure of Er$_{2}$Ti$_{2}$O$_{7}$ has AF nature,  allowing us to approximate the dispersion of the low energy acoustic spin waves with the following functional form:
\begin{equation}
\omega_c(\vec{q}) = \sqrt{ v^2 |\vec{q}|^2 + \Delta^2}, \label{eq:eqn1}
\end{equation}
where $v$ = 0.545 meV/\AA$^{-1}$ is the spin wave velocity \cite{savary2012order}, $\vec{q}$ is the reciprocal space wavevector measured from the zone center (i.e. $\vec{q} = \vec{Q} - (1,1,1)$), and $\Delta$ is the spin wave gap.  This dispersion will be broadened by the intrinsic (finite) lifetime of the spin excitations, which is often modelled by a damped harmonic oscillator. Such damped spin waves produce an imaginary susceptibility, $\chi''$,  which follows a Lorentzian profile in energy with variable full width at half max (FWHM) $\Gamma$: $L(x,\Gamma) = \frac{\frac{1}{2}\Gamma}{\pi(x^2 + (\frac{\Gamma}{2})^{2})}$.  The imaginary (i.e. absorptive) part of the susceptibility is modeled as,  


\begin{multline}
\label{eqn:S}
\chi''(\vec{q}, \omega) \propto \frac{1}{\omega_{c}(\vec{q})}  \bigg[ L\big(\omega - \omega_{c}(\vec{q}), \Gamma\big) - L\big(\omega + \omega_{c}(\vec{q}), \Gamma\big) \bigg].
\end{multline}

INS measures the dynamic structure factor, $S(\vec{q},\omega)$, which is related to this by $S(\vec{q},\omega) \propto \chi''/(1-e^{-\hbar \omega/k_B T})$.   The resolution of the measurement is broadened in energy from instrumental effects, which can be treated as a Gaussian, $R(\omega) = \frac{\exp{(-(\hbar\omega)^2/2c^2)}}{\sqrt{2\pi c^{2}}}$.  For our instrumental configuration, the resolution FWHM is $2\sqrt{(2\ln{(2))}}c$ = 0.013 meV.  The intensity is also convolved with the mosaic structure of the crystal, $M(\vec{q})$, approximated as a three-dimensional Gaussian using fixed widths, $c_H$, $c_K$ and $c_L$ , measured at the (111) Bragg peak; $M(\vec{q}) = \exp{(- (\frac{q_H^2}{c_H^2} + \frac{q_K^2}{c_K^2} + \frac{q_L^2}{c_L^2}))}$.  Here, $c_H =0.068 $\AA$^{-1}$, $c_K = 0.031 $\AA$^{-1}$.  The out-of-plane mosaic width,  $c_L$, cannot be measured using DCS and is taken to be equal to $c_H$ since they both represent widths along transverse $\vec{q}$-directions.

The expected INS intensity, with scale factor $A$, is
\begin{multline}
\label{eqn:scalc}
I(\vec{q},\omega)= A \iint S(\vec{q'},\omega')R(\omega-\omega')M(\vec{q} - \vec{q'}) d\omega'd^{3}q'. 
\end{multline}

The best fit description of this model to the inelastic scattering at T=60 mK is shown in Fig. 1 b), where it can be directly compared to the measurement in Fig. 1 a).  The related comparison between
the calculated and measured $S(\vec{Q}, \omega)$ at T= 700 mK is made in Fig. 1 d) and c), respectively.  Clearly this model provides an excellent description of the inelastic scattering near the (1,1,1) magnetic zone center, and it yields a base temperature ObQD spin wave gap of $\Delta$=0.053 $\pm$ 0.006 meV.

The temperature dependence of the ObQD gap can be determined by fits to intensity vs energy scans, approximating constant-$\vec{Q}$ scans at the (1,1,1) magnetic zone center, as shown in Fig. 2 a).  The inelastic scattering here has been integrated over the region 0.30 $\le$ H $\le$ 0.36 r.l.u. and 0.58 $\le$ K $\le$ 0.74 r.l.u., as indicated within the red rectangle around (1,1,1) shown in Fig. 2 b).  The resulting fits to the data, using a numerical evaluation of Eq. 3, are shown as the red solid lines overlaid on the inelastic data in Fig. 2 a).  The description of the data is clearly excellent, and the temperature dependence of the three parameters within Eq. 3 extracted from the fits, the ObQD gap, $\Delta$, the intrinsic FWHM, $\Gamma$, and the scale factor, $A$, are shown in Fig. 2 c).  The ObQD gap, $\Delta$ in the top panel of Fig. 2 c), is observed to decrease from its base temperature value of $\Delta$=0.053 $\pm$ 0.006 meV in what appears to be a continuous fashion.  This same temperature dependence is seen in the elastic magnetic scattering at (1,1,1), while the scale factor $A$ remains constant (bottom of Fig. 2 c).  The energy width of the inelastic peak, $\Gamma$, is shown in the middle panel of Fig. 2 c), and it stays roughly constant at ~0.04 meV (nearly three times the instrumental resolution, shown as a red line).  The reason for the finite intrinsic energy width of the zone center spin wave is not clear, but it may be related to magnons interacting with domain walls between the six degenerate $\psi_2$ domains \cite{ruff2008, de2012magnetic, savary2012order}.                                                              

We can make a more detailed comparison between the temperature dependence of the ObQD gap, $\Delta$, and the magnetic order parameter in Er$_2$Ti$_2$O$_7$, and this is what is shown in Fig. 3.  Here we overlay the temperature dependence of $\Delta$, with the measured magnetic Bragg intensity at the (1,1,1) elastic position in this experiment, and the previously measured magnetic intensity at the (2,2,0) elastic position \cite{champion2003}.  The two sets of elastic Bragg scattering are normalized to each other at 700 mK.  As the magnetic Bragg intensity varies as $M(\vec{Q})^2$, Fig. 3 shows the ObQD gap, $\Delta$, to also vary as $M(\vec{Q})^2$.

To conclude, inelastic magnetic neutron spectroscopy reveals a gap in the spin wave spectrum of Er$_2$Ti$_2$O$_7$ of $\Delta$=0.053 $\pm$ 0.006 meV at the magnetic zone center.  This determination is consistent with upper limits placed on it by analysis of low temperature heat capacity, and with theoretical expectations based on a robust microscopic spin Hamiltonian.  Previous theoretical work shows that such a gap cannot originate from small energetic terms in the Hamiltonian.  Rather the gap is induced by fluctuations.  It is an important and defining characteristic of the ObQD mechanism for ground state selection in Er$_2$Ti$_2$O$_7$, now comprehensively established. 

This work utilized facilities supported in part by the National Science Foundation under Agreement No. DMR-0944772.  Work at McMaster University was supported by NSERC of Canada.  The authors are grateful to L. Savary and L. Balents for discussions.  K.A.R. acknowledges helpful comments from C.L. Broholm and M. Mourigal.

%

\end{document}